\documentclass[aps,twocolumn]{revtex4}
\usepackage{amsmath,amssymb}
\usepackage{mathrsfs}

\begin{document}
\title{Projector operators for the no-core shell model}
\author{A. M. Shirokov}
\affiliation{Skobeltsyn Institute of Nuclear Physics, Moscow State University,
Moscow, 119992, Russia}

\begin{abstract}
Projection operators for the use within {\itshape ab initio}  no-core
  shell model, are suggested. 
\end{abstract}
\maketitle

\section{Introduction}
Shell model is a recognized tool for microscopic studies of nuclear
structure. No-core shell model (NCSM) \cite{Vary,Vary2,Vary3}, the
version of the shell model where all $A$ nucleons are
spectroscopically active,  is widely used now (see, e.~g.,
Refs. \cite{Vary3,NaO,ISTP,JISP6}) for {\it ab initio} calculations of
light nuclei (up through $A=16$) with modern realistic nucleon-nucleon
and three-nucleon forces. 

NCSM utilizes the basis of Slater determinants of single-particle
oscillator states. These basis functions are known to have spurious
contributions of center-of-mass (CM) excitations. The 
wave functions of physically
acceptable eigenstates of intrinsic NCSM Hamiltonian
\begin{equation}
\label{intrinsic}
H_A=\frac{1}{A}\sum_{i<j}^A\frac{({\bf p}_i-{\bf p}_j)^2}{2m}
+\sum_{i<j}^A V_{NN,ij}+\!\sum_{i<j<k}^A\!\! V_{NNN,ijk},
\end{equation}
where $m$ is the nucleon mass, $V_{NN,ij}$ is the two-nucleon
 interaction (including both strong and electromagnetic components),
 $V_{NNN,ijk}$ is the three-nucleon  interaction,
 should be arranged as spurious-free linear
combinations of basis states. 

To achieve this, the auxiliary 
Hamiltonian 
\begin{equation}
\label{realH}
H_{NCSM}=H_A+\beta\widetilde{Q}_0
\end{equation}
is  conventionally diagonalized within NCSM instead of the Hamiltonian
(\ref{intrinsic}). Here
\begin{gather}
\widetilde{Q}_0\equiv H_{CM}-\frac32\hbar\Omega,
\label{til-Q}  \\
\label{HCM}
H_{CM}=T_{CM}+U_{CM}
\end{gather}
is the harmonic oscillator CM Hamiltonian,
$T_{CM}$ is the CM kinetic energy operator, and
\begin{gather}
\label{UCM}
U_{CM}=\frac12 Am\Omega^2{\bf R}^2,
\intertext{where }
\label{RCM}
{\bf R}=\frac{1}{A} \sum_{i=1}^A{\bf r}_i.
\end{gather}
The term $\beta\widetilde{Q}_0$
with large
enough parameter $\beta$ has no effect
on the intrinsic states of the $A$-body system due to the
translational invariance of the Hamiltonian (\ref{intrinsic}), it
shifts up in energy spurious CM-excited states and projects out the
spurious contributions in the
low-lying eigenstates. As a result, the physical low-lying eigenstates
of (\ref{realH}) correspond to the $0\hbar\Omega$ CM-excitation and
are independent on the choice of~$\beta$.  

I suggest below a projection operator $P_{CM}$ that can be used to
project out spurious CM-excited components and to obtain spurious-free
linear combinations of basis Slater determinants that can be used as a
new spurious-free basis for direct diagonalization of the
intrinsic Hamiltonian (\ref{intrinsic}). The  complete spurious-free basis 
corresponding to  the $0\hbar\Omega$ CM-excitations, 
is much smaller than the basis of all
Slater determinants including all  $\varkappa\hbar\Omega$
CM-excitations with $\varkappa\leq N$ where $N$ is the maximal
oscillator quanta of the 
$N\hbar\Omega$  NCSM model space 
used in the
calculations. Therefore it is expected that the use of the  projection
operator $P_{CM}$ will simplify essentially the NCSM studies of
nuclear structure, will make it possible to arrange the calculations in a
larger $N\hbar\Omega$ 
model spaces with the same computer
facilities and hence to improve the accuracy of the NCSM predictions, etc.

I note also that the so-called $m$-scheme is conventionally utilized
in the NCSM
applications, i.~e. the basis Slater determinants are used that do not have 
definite values of the orbital angular momentum $L$, of the total
angular momentum $J$, and of the total spin $S$. 
The $m$-scheme makes it possible to  use well-developed 
computational methods and available respective computer codes. However
the basis of the  $m$-scheme 
Slater determinants is very large since it includes all states with
all possible values of $J\leq J_{\max}$,   $L\leq L_{\max}$ and
$S\leq S_{\max}$ where the maximal values  $J_{\max}$,   $L_{\max}$ and
$S_{\max}$ are large enough in modern NCSM applications and
depend on the particular nucleus under consideration and on the 
$N\hbar\Omega$ model space used in the
calculations.

 I suggest below the
projection operators $P_J$, $P_L$ and $P_S$ on the states with
definite $J$, $L$ and $S$ values. These projection operators as well
as $P_{CM}$ can be
easily utilized within the existing NCSM codes to reduce essentially
the number of the basis states.

\section{
CM-projector $P_{CM}$}
Let 
\begin{gather}
\label{Psi0Psi1}
\Psi=\sum_{\varkappa=0}^{N}\alpha_\varkappa\Psi_\varkappa
\end{gather} 
be a vector (wave function) defined in the
$N\hbar\Omega$ 
model space; $N$ is the maximal possible CM-excitation quanta in
this model space. Equation (\ref{Psi0Psi1}) presents expansion of
$\Psi$ in the series of functions $\Psi_\varkappa$ 
with a definite  CM-excitation quanta $\varkappa=0$, 1,~..., $N$. The
functions  $\Psi_\varkappa$ are the eigenfunctions of the harmonic oscillator
CM Hamiltonian:
\begin{gather}
\label{HcmPsikappa}
H_{CM}\Psi_\varkappa=\left(\varkappa+\frac32\right)\hbar\Omega\,\Psi_\varkappa.
\end{gather}

Due to Eq.~(\ref{HcmPsikappa}),
the operator $\widetilde{Q}_0$ defined by Eq.~(\ref{til-Q}) acts as
 `anti-projector': it projects out the spurious-free  component   
${\Psi}_{sf}\equiv\Psi_0$
of the wave function,
\begin{gather}
\label{Q0Psi0}
\widetilde{Q}_0{\Psi}_{sf}=0.
\end{gather}
 We can also define anti-projectors 
\begin{gather}
\widetilde{Q}_\varkappa\equiv H_{CM}-\left(\varkappa+\frac32\right)\hbar\Omega
\label{til-QN}
\end{gather}
which project out components with  given values of the CM 
excitation quanta 
$\varkappa$:  
\begin{gather}
\label{QNPsiN}
\widetilde{Q}_\varkappa{\Psi}_{\varkappa}=0.
\end{gather}

 We can extract the spurious-free content  
$\widetilde{\Psi}_{sf}$
of  $\Psi$ by the subsequent use of the  operators (\ref{til-QN}):
\begin{subequations}
\label{subsec}
\begin{gather}
\Psi_1=\widetilde{Q}_1 \Psi, \\
\Psi_2=\widetilde{Q}_2 \Psi_1, \\
 ... \notag \\
\widetilde{\Psi}_{sf}\equiv\Psi_{N}
       =\widetilde{Q}_{N} \Psi_{N-1}.
\end{gather}
\end{subequations}
Equations (\ref{subsec}) are equivalent to the following equation:
\begin{gather}
\widetilde{\Psi}_{sf} = \widetilde{P} \Psi ,
\label{quasi-pr}
\end{gather}
where the operator 
\begin{gather}
\widetilde{P}=\prod_{\varkappa=1}^{N}\widetilde{Q}_\varkappa.
\label{til-P}
\end{gather}

Let us call $\widetilde{P}$ `quasi-projector'. Mathematically $\widetilde{P}$
is not a projection operator since it does not fit the  standard
property of the projection operators,
\begin{gather}
P^2=P.
\label{P2-P}
\end{gather}
The function $\Psi$ is a superposition (\ref{Psi0Psi1}) of the 
spurious-free $\Psi_{sf}\equiv\Psi_0$ and spurious 
 components $\Psi_\varkappa$ with $\varkappa\ne 0$.
The standard projection operator property (\ref{P2-P}) guaranteers that
\begin{gather}
P\Psi=\alpha_0\Psi_{sf}.
\label{P2-P-mean}
\end{gather}
Instead of (\ref{P2-P-mean}),
the quasi-projector
$\widetilde{P}$ when applied to $\Psi$ results in 
\begin{gather}
\widetilde{P}  \Psi =\widetilde{\Psi}_{sf}   =D\alpha_0\Psi_{sf}.
\label{P2-tilP-mean}
\end{gather}
The constant $D$ can be easily calculated using
Eqs. (\ref{HcmPsikappa}) and (\ref{til-QN}):
\begin{gather}
D= (-1)^{N} {N}!\,(\hbar\Omega)^{N}.
\label{D-Prod}
\end{gather}
To become a projector, the quasi-projector  $\widetilde{P}$ should
be `normalized':
\begin{gather}
P=\frac{1}{D}\widetilde{P}.
\label{P-til-P}
\end{gather}

In applications, one can use either $\widetilde{P}$ or $P$. Really, 
it is usually needed to extract from $\Psi$ its normalized
 spurious-free component 
 $\Psi_{sf}$. The multiplier $\alpha_0$ is usually unknown. Hence after using
either the quasi-projector (\ref{til-P}) 
or the projector  (\ref{P-til-P}), one needs to normalize either the 
function $D\alpha_0\Psi_{sf}$ or the function
$\alpha_0\Psi_{sf}$.  Clearly, the same computational efforts are required to
normalize the functions $D\alpha_0\Psi_{sf}$ and~$\alpha_0\Psi_{sf}$.

\section{Other useful projectors}

The same idea can be utilized for the construction of other useful
projectors. As an example, let us construct the projector on the states
with a definite value of the angular momentum.

Let $\hat{L}^2=\hat{L}^2_x+\hat{L}^2_y+\hat{L}^2_z$  be the standard
orbital momentum operator. Its eigenvalues are known to be $L(L+1)$. 
We define now the operators
\begin{gather}
\widetilde{Q}_L\equiv \hat{L}^2 - L(L+1)
\label{til-QL}
\intertext{and}
\widetilde{P}^{L_{\max}}_L=\prod_{\kappa=0}^{L-1}\widetilde{Q}_\kappa
\prod_{\kappa=L+1}^{L_{\max}}\widetilde{Q}_\kappa,
\label{til-PL}
\end{gather}
where $L_{\max}$ is the maximal accessible orbital momentum  in the
given $N\hbar\Omega$ shell model space. The non-normalized quasi-projector
(\ref{til-PL}) can be used like the CM non-normalized quasi-projector 
(\ref{til-P}) to extract (non-normalized) component with the definite 
value of the orbital momentum $L$ by the algorithm described briefly by Eq.
(\ref{quasi-pr}) or in more detail by Eqs. (\ref{subsec}).

The  projector 
$P^{L_{\max}}_L$ can be expressed as
\begin{gather}
P^{L_{\max}}_L=\frac{1}{D^{L_{\max}}_L}\widetilde{P}^{L_{\max}}_L
\label{PL}
\end{gather}
where
\begin{multline}
D^{L_{\max}}_L=\prod_{\kappa=0}^{L-1}[\kappa(\kappa+1)-L(L+1)]\\
\times
\prod_{\kappa=L+1}^{L_{\max}}[\kappa(\kappa+1)-L(L+1)].
\label{DL}
\end{multline}

The structure of the projectors $P^{J_{\max}}_J$,  $P^{S_{\max}}_S$ 
and $P^{T_{\max}}_T$
on the states with given values of the total angular momentum $J$,
total spin $S$ or isospin $T$,
is exactly the same. The only difference is that in the case of
an odd-$A$ system, one should use half-integer  $J$, $T$ or $S$ values and
modify respectively the products in Eqs.~(\ref{til-PL}) and (\ref{DL}).

The standard projection operator property (\ref{P2-P}) is valid for
all projectors (but not quasi-projectors) discussed above. 

\section{Conclusions}

Expression of the projection operators on the states with definite
value of the angular momentum in the form of the 
expansion in the
powers of the SU(2) generators, are known in the literature (see,
e.~g., Ref.~\cite{FOS}). However, in the general case, this polynomial
includes an infinite number of terms and is inconvenient for the use
in the
nuclear shell model applications. As it was shown above, in the case
of the shell model, the projector can be taken in the form of a finite
polynomial in generators that is much more useful for the
applications. 
The suggested projectors
$P^{L_{\max}}_L$,  $P^{J_{\max}}_J$,  $P^{S_{\max}}_S$ and
$P^{T_{\max}}_T$ are of this form.

The CM-projector $P_{CM}$ is also suggested as a finite expansion in
the powers of a simple CM harmonic oscillator operator $H_{CM}$. To the
best of my knowledge, similar expression for the CM-projector were
never discussed in the literature.

The Lanczos iteration approach is utilized in the modern shell model
codes, i.~e. the basis vectors are obtained successively by acting by
the Hamiltonian  on the vector obtained on the previous step. The
intrinsic Hamiltonian (\ref{intrinsic}) and the NCSM Hamiltonian (\ref{realH})
cannot produce CM-excited
states or to change the value of the total angular momentum of the
state. Hence it is possible to project only the pivot vector
(the initial vector in the  Lanczos iteration approach) on the
spurious-free subspace with the given definite value of the total
angular momentum $J$; all the rest basis vectors will be produced 
spurious-free and with the same value of $J$ by
the Lanczos iterations. 

Formally one can use the projected pivot vector and the intrinsic
Hamiltonian (\ref{intrinsic}) instead of the auxiliary Hamiltonian
(\ref{realH}) in the NCSM applications. However it is well known that
the spurious states will be produced in the  Lanczos iteration
approach due to the computer noise (round-off errors). The term
$\beta\widetilde{Q}_0$ in Eq.~(\ref{realH}) stabilizes the NCSM
calculations reducing the computer noise if $\beta$ is sufficiently
large. Therefore it looks reasonable to utilize the auxiliary
Hamiltonian (\ref{intrinsic}) in the applications; probably it is
reasonable to add the term 
$\gamma |\hat J^2 - J(J+1)|$ with  sufficiently large $\gamma$ to the Hamiltonian
(\ref{realH}) to reduce the computer noise in the calculations of the
states with the definite value $J$ of the total angular momentum.

\mbox{}\\

The author is thankful to Yu.~F.~Smirnov, J.~P.~Vary and P.~Navr\'atil
for fruitful discussions.
This work was supported in part
by the Russian Foundation of Basic Research  grant No 05-02-17429.


\end{document}